\documentclass[prl,cha,twocolumn,superscriptaddress,preprintnumbers,showpacs, amsmath,amssymb]{revtex4-1}
\usepackage{bm}
\usepackage{ulem}
\usepackage[colorlinks=true,linkcolor=blue,citecolor=blue]{hyperref}
\usepackage{amsmath}
\usepackage{amssymb}
\usepackage{amsthm}
\usepackage{amsfonts}
\usepackage{enumerate}
\usepackage{latexsym}
\usepackage{ifpdf}
\usepackage{graphicx}
\usepackage{makeidx}
\expandafter\ifx\csname package@font\endcsname\relax\else
\expandafter\expandafter
\expandafter\usepackage
\expandafter\expandafter
\expandafter{\csname package@font\endcsname}
\fi
\begin{document}

\title[]{Creating and manipulating interfacial spin with giant magnetic response in 4$f$ antiferromagnets}
\author{Ruyi Zhang}
\thanks{These authors contributed equally to this work}
\affiliation{Ningbo Institute of Materials Technology and Engineering, Chinese Academy of Sciences, Ningbo, Zhejiang 315201, China.}
\affiliation{Center of Materials Science and Optoelectronics Engineering, University of Chinese Academy of Sciences, Beijing 100049, China.}
\author{Yujuan Pei}
\thanks{These authors contributed equally to this work}
\affiliation{Ningbo Institute of Materials Technology and Engineering, Chinese Academy of Sciences, Ningbo, Zhejiang 315201, China.}
\affiliation{Center of Materials Science and Optoelectronics Engineering, University of Chinese Academy of Sciences, Beijing 100049, China.}
\affiliation{State Key Laboratory of Metastable Materials Science and Technology, Yanshan University, Qinhuangdao 066004, China}
\author{Yang Song}
\affiliation{Ningbo Institute of Materials Technology and Engineering, Chinese Academy of Sciences, Ningbo, Zhejiang 315201, China.}
\affiliation{Center of Materials Science and Optoelectronics Engineering, University of Chinese Academy of Sciences, Beijing 100049, China.}
\author{Jiachang Bi}
\affiliation{Ningbo Institute of Materials Technology and Engineering, Chinese Academy of Sciences, Ningbo, Zhejiang 315201, China.}
\affiliation{Center of Materials Science and Optoelectronics Engineering, University of Chinese Academy of Sciences, Beijing 100049, China.}
\author{Jingkai Yang}
\affiliation{State Key Laboratory of Metastable Materials Science and Technology, Yanshan University, Qinhuangdao 066004, China}
\author{Junxi Duan}
\affiliation{School of Physics, Beijing Institute of Technology, Beijing 100081, China}
\author{Yanwei Cao}
\email{ywcao@nimte.ac.cn}
\affiliation{Ningbo Institute of Materials Technology and Engineering, Chinese Academy of Sciences, Ningbo, Zhejiang 315201, China.}
\affiliation{Center of Materials Science and Optoelectronics Engineering, University of Chinese Academy of Sciences, Beijing 100049, China.}
\date{\today}

\begin{abstract}
Creating and manipulating spin polarization in low-dimensional electron systems (such as two-dimensional electron gases) is fundamentally essential for spintronic applications, which is yet a challenge to date. In this work, we establish the metamagnetic phase diagram of 4$f$ antiferromagnetic TbScO$_3$ and reveal its giant magnetic response to sub-tesla magnetic field, which has not been reported thus far. Utilizing this giant magnetic response, we demonstrate that the spin polarization of two-dimensional electron gas in SrTiO$_3$/LaTiO$_3$/TbScO$_3$ heterostructure can be manipulated successfully in aid of interfacial 3\textit{d}-4\textit{f} exchange interaction. Remarkably, the hysteretic magnetoresistances of two-dimensional electron gas at the SrTiO$_3$/LaTiO$_3$ interface are entirely determined by the metamagnetic phase transitions of the underlying TbScO$_3$ substrate. Our results pave a novel route to engineer the spin polarization of low-dimensional electron systems in 4$f$ antiferromagnet-based heterostructures.
\end{abstract}

\maketitle
Creating, detecting and manipulating spin degrees of freedom is at the center of modern condensed matter physics and spintronics, the study of which is fundamentally important not only for basic science but also for device applications \cite{Fert-RMP-2008, Bader-ARCMP-2010, Baltz-RMP-2018}, e.g., the key roles of spin in topological states \cite{Sessi-Science-2016}, skyrmions \cite{Yu-Nature-2018}, antiferromagnetic (AFM) devices \cite{Chen-NM-2019}, and spin field-effect-transistors \cite{Koo-Science-2009}. Among them, due to the great potential for future spin field-effect-transistor application, the study of magnetic semiconductors and spin-polarized two-dimensional electron gases (SP2DEGs) in semiconductor heterostructures and oxide interfaces have attracted tremendous interests \cite{Lee-NM-2013, Zhang-PRL-2018}. To create SP2DEGs, several routes have been developed such as defects/impurities doping \cite{Lee-NM-2013}, spin injection \cite{Ohshima-NM-2017,Duan-APL-2013}, ferroelectric polar gating \cite{Zhang-NC-2018}, interfacial charge transfer \cite{Cao-PRL-2016}, and magnetic proximity effect \cite{Kareev-APL-2013}. To manipulate the spin polarization, methods such as spin-polarized current, electric field, and photonic field have been exploited \cite{Bader-ARCMP-2010, Stornaiuolo-NM-2016, Nemec-NP-2018}. Very recently, deep understanding and control of AFM order promote the rise of AFM spintronics \cite{Baltz-RMP-2018, Zelezny-NP-2018, Lebrun-Nature-2018}. It has been reported that the magnetic response of antiferromagnets to temperature, electric field, and pressure can lead to many nontrivial magnetism-related properties including multiferroicity \cite{Lebeugle-PRL-2008}, spin Hall effect \cite{Cheng-PRL-2016}, and superconductivity \cite{Manna-NC-2017}. However, due to locked pairs of spin, the AFM order is insensitive to the magnetic field unless it is very large \cite{Bluschke-PRL-2017, Hao-NP-2018}. Therefore, it raises an interesting question that whether the spin polarization of 2DEG could be manipulated by the magnetic response in antiferromagnetic materials or not.

\begin{figure}[t]
\includegraphics[width=0.4\textwidth]{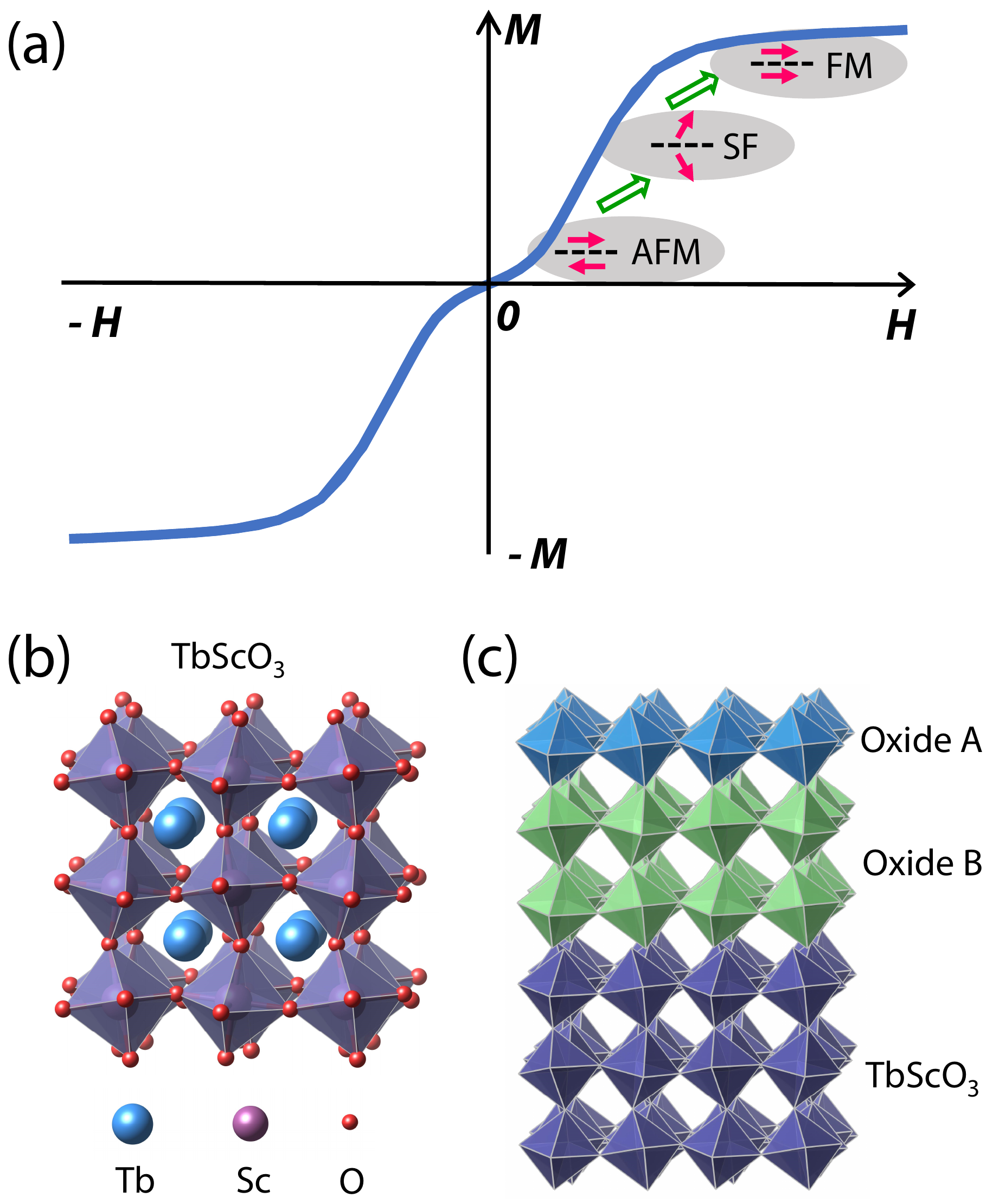}
\caption{ \label{Fig1}(a) Schematic of giant response to magnetic field and metamagnetic transitions in 4$f$ antiferromagnets. (b) Crystal structure of TbScO$_3$. (c) TbScO$_3$-based oxide heterostructures.}
\end{figure}

To address above, we take AFM scandate TbScO$_3$ (TSO) as a prototype to investigate the metamagnetic phase transitions and their critical roles in manipulating the spin polarization of 2DEG. The rare-earth scandates are widely used as dielectric gate materials and single crystal substrates for film deposition owing to their high dielectric constant (18 - 35.5) \cite{Delugas-PRB-2007}, large optical band gap ($>$ 4.9 eV) \cite{Derks-PRB-2012}, excellent thermal stability \cite{Adelmann-APL-2008}, and structural, chemical, and thermal compatibilities with various perovskite thin films (e.g., SrTiO$_3$ \cite{Haeni-Nature-2004}, BiFeO$_3$ \cite{Sando-NM-2013}, SrRuO$_3$ \cite{Wei-NC-2017}, etc.). As a member of rare-earth scandates, TSO possesses Sc$^{3+}$ cations in diamagnetic state and Tb$^{3+}$ cations with 4\textit{f} electrons governing the magnetism. Despite its wide applications, the magnetic properties of TSO has not been investigated in detail yet. Meanwhile, the 4\textit{f} antiferromagnets are unique with strong exchange interaction and comparatively weak magnetic anisotropy, which show unusual response to magnetic field (see Fig. \ref{Fig1}(a)). With increasing the magnetic field to a critical point, where Zeeman energy begins to compensate the energy of magnetic anisotropy, the antiparallel spins in 4\textit{f} antiferromagnets prefer to rotate freely to the direction perpendicular to the magnetic easy axis to minimize the magnetostatic energy \cite{Baltz-RMP-2018}. Generally, this field induced spin reorientation phenomenon is named spin flop (SF) transition, a metamagnetic transition \cite{Baltz-RMP-2018, Machado-PRB-2017}. Furthermore, owing to the transfer of magnetic order and anisotropy through interfacial 3$d$-4$f$ exchange interaction \cite{Bluschke-PRL-2017}, it paves a possible way to manipulate the spin polarization of 2DEG in TSO-based oxide heterostructure (see Fig. \ref{Fig1}(b) and \ref{Fig1}(c)). 

\begin{figure}[]
\includegraphics[width=0.5\textwidth]{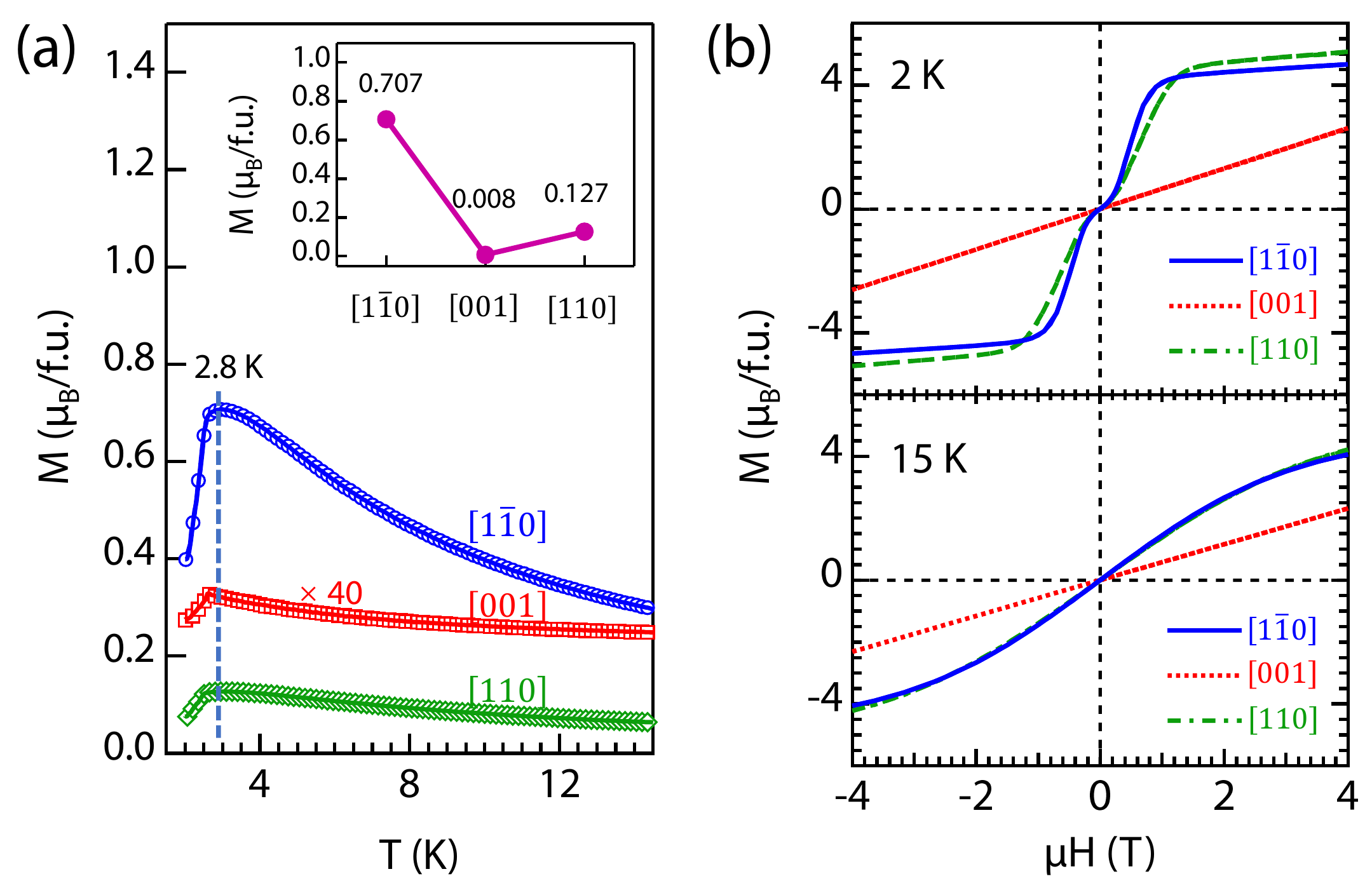}
\caption{\label{Fig2} (a) ZFC (marker) and FC (solid line) measurements with $\mu$H = 0.2 T along [1$\bar{1}$0] (blue), [001] (red), and [110] (green) directions. The amplitude of magnetization along [001] is multiplied by 40 for better view. Inset, orientation-dependent magnetic moments near 2.8 K with $\mu$H = 0.2 T. (b) Orientation-dependent M-H curves at 2 K and 15 K, respectively.}
\end{figure}

In this letter, we demonstrate the spin polarization of 2DEG can be effectively manipulated by the metamagnetic phase transitions in AFM TSO. We reveal the phase transitions from AFM to SF, and then to ferromagnetic (FM) states with sub-tesla magnetic fields, which are significantly lower than the fields required in many other AFM materials. Particularly, in the heterostructure SrTiO$_3$/LaTiO$_3$/TbScO$_3$, we show that the spin polarization of 2DEG at SrTiO$_3$/LaTiO$_3$ interface can be completely manipulated by the metamagnetic phase transitions of TSO in aid of interfacial 3\textit{d}-4\textit{f} magnetic exchange interaction. Our study indicates TSO could be a great platform to create and manipulate the spin polarization in 4$f$ antiferromagnet-based heterostructures.

Bulk TSO has an orthorhombic perovskite structure with space group \textit{Pbnm} and lattice parameters a=5.466 \AA, b=5.731 \AA, c=7.917 \AA \cite{Uecker-JCG-2008}. The low temperature magnetism ranging from 2 K to 15 K of TSO(110) single crystal was measured using a Quantum Design MPMS superconducting quantum interference device (SQUID). The zero-field-cooled (ZFC) and field-cooled (FC) measurements along three main crystallographic orientations [1$\bar{1}$0], [110], and [001] (in orthorhombic annotations) were recorded with applied magnetic field ranging from 0.005 T to 2 T. The magnetic hysteresis (M-H) curves with magnetic fields ramping between $\pm$ 7 T were recorded with temperature varying from 2 K to 15 K.

Firstly, we investigate the anisotropic antiferromagnetism of TSO, which highly depends on the interactions among different magnetic parameters (e.g., exchange interaction, magnetic anisotropy, and Zeeman energy) \cite{Li-NCM-2016}. As seen in Fig. \ref{Fig2}(a), the dominative feature in ZFC/FC measurement is the presence of critical temperature $\sim$ 2.8 K, indicating an AFM to paramagnetic (PM) transition upon warming. It is noted that the AFM order is strongly anisotropic, e.g., the maximum magnetic moments (with 0.2 T magnetic field) are 0.707, 0.008, and 0.127 $\mu$$_B$ along [1$\bar{1}$0], [001], and [110] directions, respectively (see inset in Fig. \ref{Fig2}(a)), indicating the easy axis and hard axis are more inclined to [1$\bar{1}$0] and [001] directions, respectively. The strong magnetic anisotropy can be further verified by the M-H curves shown in Fig. \ref{Fig2}(b). As seen, the feature of SF transition is apparent along both [1$\bar{1}$0] and [110] directions at 2 K, whereas it is absent along [001] orientation. With further increasing the temperature to 15 K, the SF transition is totally suppressed.

\begin{figure}[]
\includegraphics[width=0.4\textwidth]{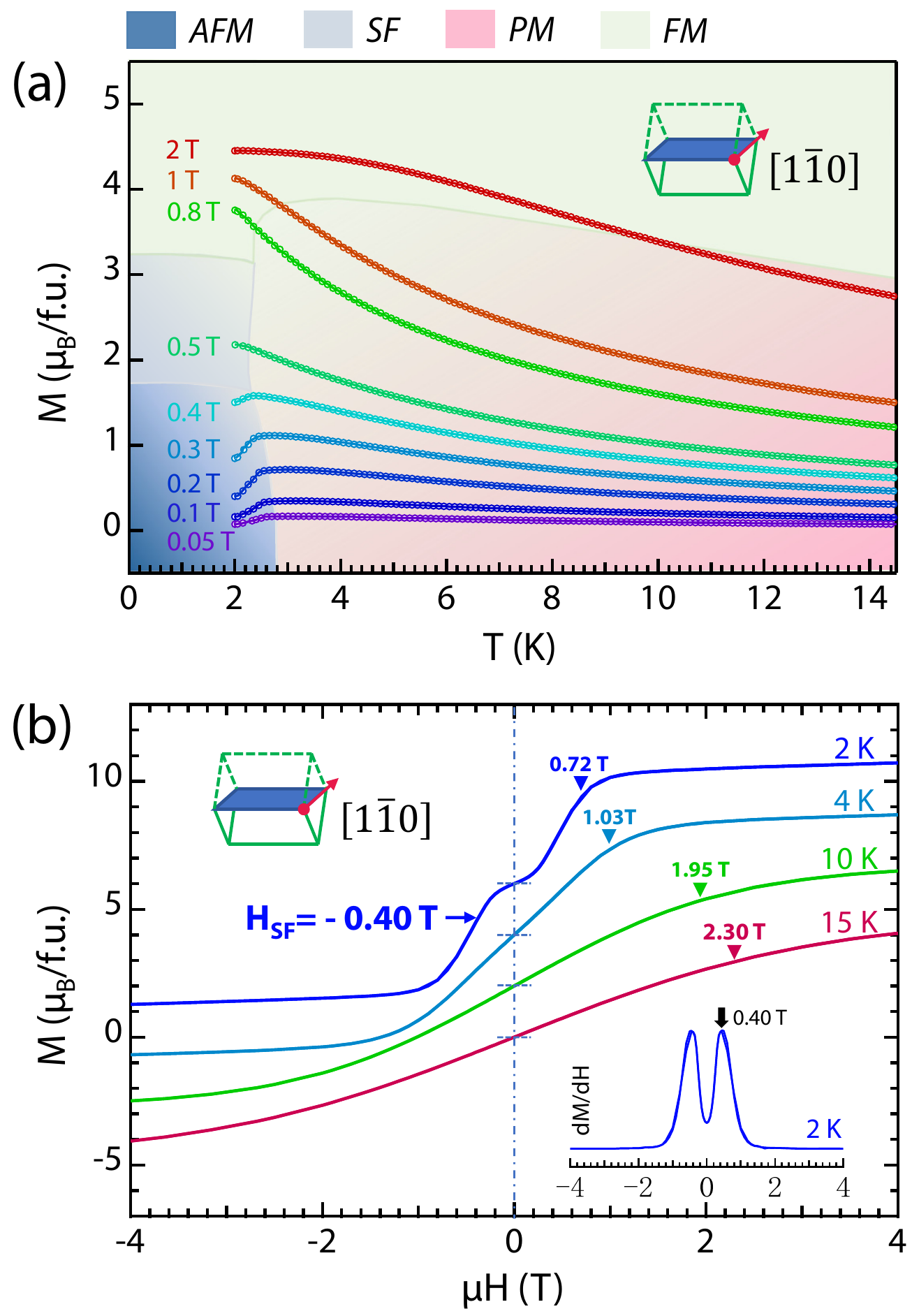}
\caption{\label{Fig3} Field- and temperature-dependent magnetizations measured along [1$\bar{1}$0] direction. (a) Field-dependent ZFC (circles) and FC (solid lines) measurements. (b) M-H curves from 2 K to 15 K, which are vertically shift for clarity. The SF-FM and PM-FM transition fields are marked by triangles. The inset shows dM/dH curve at 2 K.}
\end{figure}

Next, to establish the phase diagram of metamagnetic phase transition in TSO, we measure the magnetic field- and temperature-dependent magnetizations along [1$\bar{1}$0] direction (corresponding to in-plane [100] direction for pseudo-cubic geometry) \cite{Sando-NM-2013,Wei-NC-2017}. As seen in Fig. \ref{Fig3}(a), it shows AFM-PM transitions at Néel temperature \textit{T$_N$} $\sim$ 3.0 K with small magnetic fields $\leq$ 0.05 T, whereas T$_N$ drops with increasing magnetic field, indicating the melting of AFM order due to enhanced Zeeman interaction \cite{Ke-APL-2009}. As field further increases, the Zeeman energy surpasses magnetic anisotropy, driving TSO into SF state with canted spins \cite{Baltz-RMP-2018}. With field continuously increasing, the Zeeman energy finally prevails AFM exchange interaction and forces all spins to align along the direction of magnetic field, leading to the long-range ordered FM state. The critical parameters of SF-PM and FM-PM phase transitions can be extracted from dM/dT curves \cite{Emmanouilidou-JMMM-2019,Huang-PRB-2011}. To further understand the metamagnetic phase transitions, we also measure temperature-dependent M-H curves (see Fig. \ref{Fig3}(b) and Fig. S1 in supplementary material). As seen, the M-H curve at 2 K presents a steep upturn near $\pm$ 0.4 T, which is a typical feature for SF transition \cite{Baltz-RMP-2018, Machado-PRB-2017}. The upturn disappears above 2.5 K indicating the presence of SF-PM transition (see Fig. S1(b) in supplementary material). At slightly higher magnetic field, TSO undergoes SF-FM transitions below 2.5 K and PM-FM transitions above 2.5 K. Here the critical parameters for AFM-SF and PM-FM (including SF-FM below 2.5 K) transitions are determined by the peak of dM/dH \cite{Machado-PRB-2017, Huang-PRB-2011} and the valley of d$^2$M/dH$^2$ \cite{Nilsen-PRB-2017} curves, respectively (see Fig. S2 in supplementary material). On the other hand, it is indicated that the PM-FM (including SF-FM) transition fields increase upon warming, suggesting larger Zeeman energy is required to suppress the thermal spin fluctuations at the higher temperature. As seen in Fig. 3, the magnetic moment can be as large as 5.5 $\mu$$_B$, indicating a huge magnetic response compared to typical oxide magnets \cite{Wei-NC-2017,Tsui-APL-2000,Liu-APL-2014}.

\begin{figure*}[]
\includegraphics[width=0.7\textwidth]{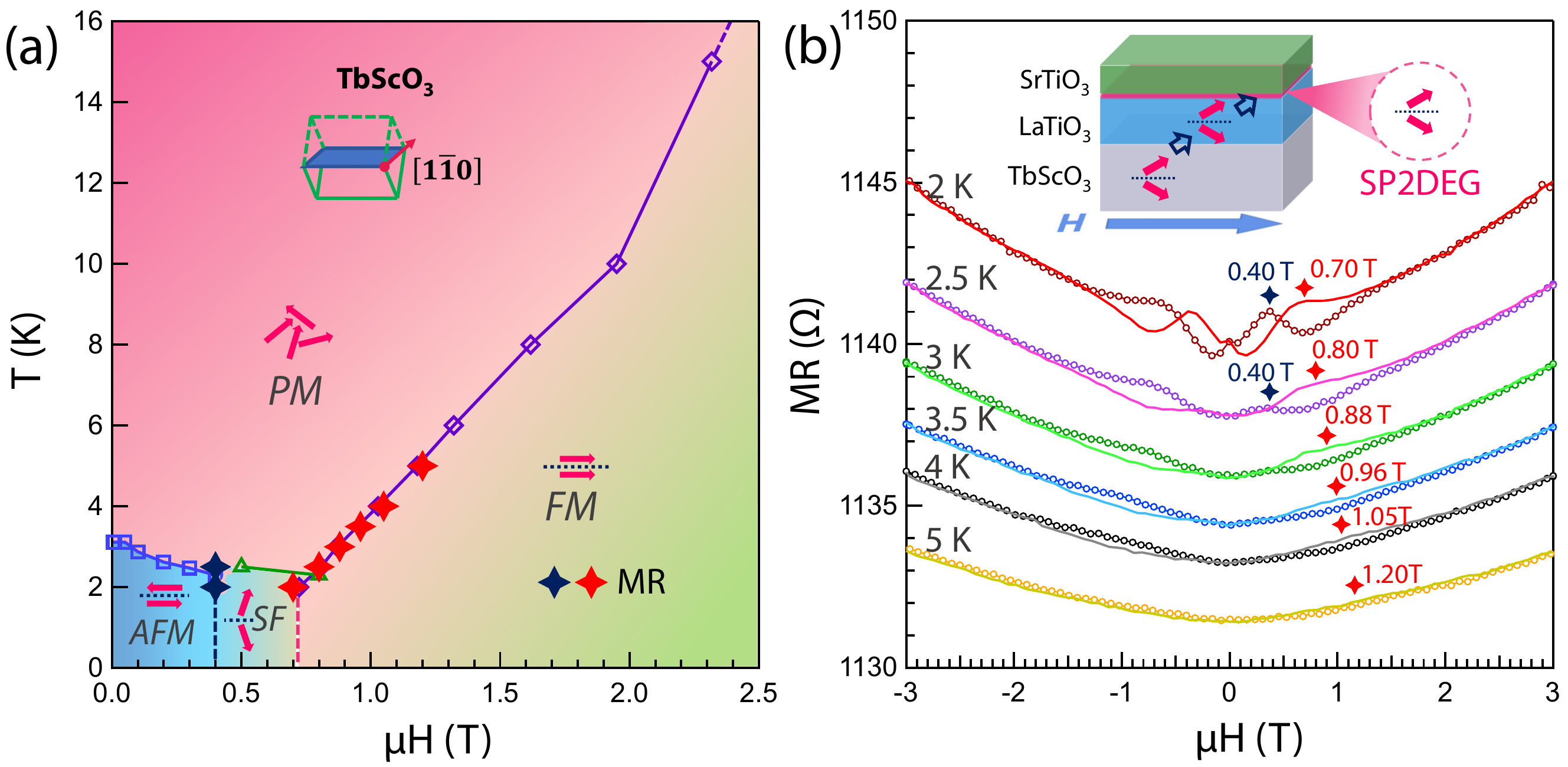}
\caption{\label{Fig4}(a) Metamagnetic phase diagram of TSO along [1$\bar{1}$0] direction. The stars label the critical parameters extracted from the magnetoresistance curves. (b) Temperature-dependent in-plane magnetoresistances (field applied along [1$\bar{1}$0] direction) in SrTiO$_3$/LaTiO$_3$/TbScO$_3$\cite{Wen-APL-2018}. The dotted thin lines and thick lines represent forward and backward sweep directions of the external magnetic field, respectively.}
\end{figure*}

Based on the magnetic response to temperature and external magnetic field, we can establish the metamagnetic phase diagram and discuss its key role in manipulating the spin polarization of 2DEG in SrTiO$_3$/LaTiO$_3$/TbScO$_3$ (STO/LTO/TSO) \cite{Wen-APL-2018}. As seen in Fig. \ref{Fig4}(a), there are four magnetic phases, which are AFM (low T and low H), SF (low T and intermediate H), FM (high T and high H), as well as PM (high T and low H), respectively. Remarkably, the field-driven FM state can survive at temperature $\sim$ 15 K, which is well above AFM temperature $\sim$ 2.8 K. More importantly, it is revealed that the critical parameters (blue and red stars in Fig. 4(b)) extracted from the magnetoresistances (MR) of SP2DEG in STO/LTO/TSO agree very well with the phase transitions in TSO (see Fig. S2 in supplementary material), demonstrating the spin polarization of 2DEG in the STO/LTO interface can be created and manipulated by the giant magnetic response in TSO. It is noted that the understanding of temperature- and magnetic field-dependent oxide SP2DEGs has been a challenge for a long time, which can neither be perfectly explained by magnetic phase separations \cite{Ayino-PRM-2018, Yee-PRX-2015} nor by spin-flip scattering process \cite{Mehta-NC-2012}. 

Next, we discuss the mechanism of creating and manipulating spin polarization of 2DEG in STO/LTO/TSO via giant magnetic response in TSO layer. It has been well-known that LTO is a G-type AFM Mott insulator with $3d^{1}$ electronic configuration \cite{Khaliullin-PRL-2000,Fritsch-PRB-2002}. The canted AFM order in LTO can survive below 146 K. Based on the model of magnetic transfer between 3$d$ LaNiO$_3$ and 4$f$ DyScO$_3$ in LaNiO$_3$/DyScO$_3$ (LNO/DSO) interfaces \cite{Bluschke-PRL-2017}, the 3\textit{d}-4\textit{f} exchange interaction can also present in LTO/TSO interface in our work. The 3\textit{d} moments within transition metal oxide layer are pined to the directions determined by the single-ion anisotropy of the Dy$^{3+}$ cations in the LNO-DSO system \cite{Bluschke-PRL-2017}. Therefore, the metamagnetic phase transitions of 4$f$ Tb$^{3+}$ in TSO layer can be transferred to 3$d$ Ti$^{3+}$ near LTO/STO interface (see Fig. \ref{Fig4}(b)) \cite{Bluschke-PRL-2017, Zhang-PRL-2018}. It is important to note that the successful transfer of magnetic order from 4\textit{f} to 3\textit{d} moments strongly relies on the AFM order of 3$d$ layer in both LNO/DSO and STO/LNO/TSO cases, which are thin LNO (2 unit cells) and thick LTO (20 unit cells) layers, respectively \cite{Bluschke-PRL-2017,Wen-APL-2018}. One of the unique features in our STO/LTO/TSO system is that the manipulation of SP2DEG can not only work out in in-plane configuration but also in out-of-plane configuration (See Fig. S3 and S4 in supplementary material), which may be closely connected to the small magnetic anisotropy ($\sim$ 4.2 meV) of LTO \cite{Keimer-PRL-2000,Khaliullin-PRB-2001}. 

In conclusion, we establish the metamagnetic phase diagram in TSO and reveal its giant magnetic response to sub-tesla magnetic field. The Curie temperature $\sim$ 15 K of field induced ferromagnetic state is more than five times higher than the Néel temperature $\sim$ 2.8 K of antiferromagnetic state. More importantly, in STO/LTO/TSO heterostructure, we demonstrate that the spin polarization of 2DEG in STO/LTO interface can be manipulated successfully via interfacial 3\textit{d}-4\textit{f} magnetic exchange interactions between LTO and TSO layers. We highlight that the route to creating and manipulating SP2DEGs demonstrated in our work is highly desired in the field of spintronic devices \cite{Baltz-RMP-2018, Zelezny-NP-2018, Lebrun-Nature-2018}. Moreover, in contrast to extremely large magnetic fields required to trigger the magnetic response in most AFM materials (e.g., $\sim$ 5 T for NiO and 42 T for FeF$_2$) \cite{Machado-PRB-2017,Jaccarino-JMMM-1983}, sub-tesla magnetic field is sufficient to induce a giant response in TSO. Our results pave a novel route to create and manipulate the spin polarization of 4$f$ antiferromagnet-based heterostructures

See the supplementary material for the SQUID measurements as a function of temperature, MR, M-H, dM/dH, and d$^2$M/dH$^2$ curves of TSO in in-plane and out-of-plane configurations, and out-of-plane metamagnetic phase diagram.

This work is supported by the National Natural Science Foundation of China (Grant No. 11874058), the Pioneer Hundred Talents Program of Chinese Academy of Sciences, the Ningbo 3315 Innovation Team, and the Ningbo Science and Technology Bureau (Grant No. 2018B10060). This work is partially supported by China Postdoctoral Science Foundation (Grant No. 2018M642500) and Postdoctoral Science Foundation of Zhejiang Province (Grant No. zj20180048).

\end{document}